\documentclass[manuscript, 11pt]{stjour} 
\usepackage{ulem}

\usepackage{amssymb}
\usepackage{fontenc}
\usepackage{amsmath}
\usepackage{amssymb}
\usepackage{float}
\usepackage{multirow}
\usepackage{listings}
\usepackage{color}
\usepackage{ulem}
\usepackage{tabularx}
\newcolumntype{L}[1]{>{\raggedright\arraybackslash}p{#1}}
\newcolumntype{C}[1]{>{\centering\arraybackslash}p{#1}}
\newcolumntype{R}[1]{>{\raggedleft\arraybackslash}p{#1}}
\usepackage{microtype}
\usepackage{booktabs,tabularx}
\usepackage{tabu}
\usepackage{listings}

\usepackage{hyperref}
\hypersetup{colorlinks, citecolor={blue},linkcolor={black}, urlcolor={gray}, pdftitle={quantile regression}, pdfauthor={Marzieh Shahmandi}, pdfdisplaydoctitle}

\usepackage{ragged2e}
\justifying

\newcommand{\eg}{\textit{e.g.}}
\newcommand{\orcid}[1]{\href{https://orcid.org/#1}{\includegraphics[width=8pt]{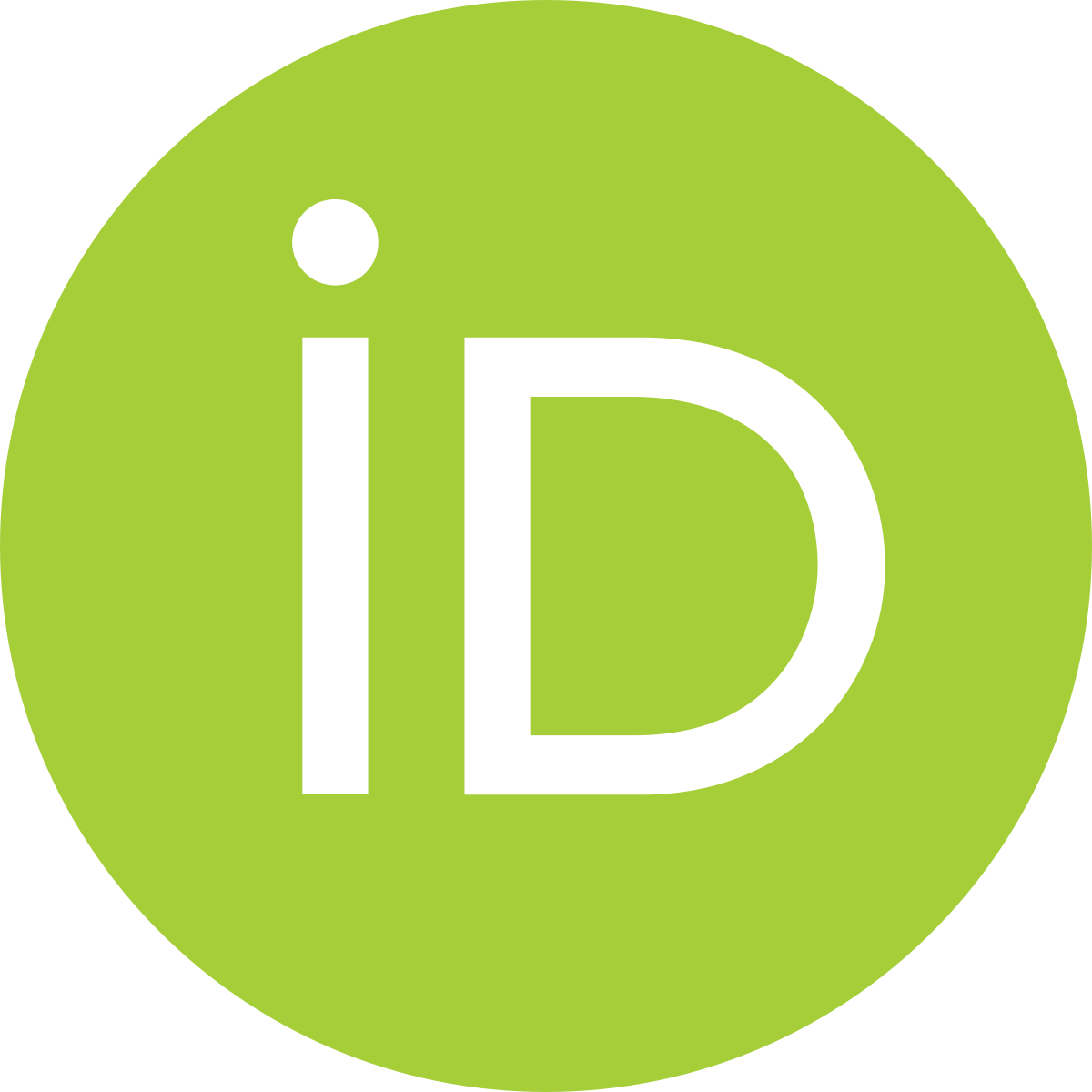}}}

\nolinenumbers
\begin{document}

\title[Bayesian Hurdle Quantile Regression]{A Bayesian Hurdle Quantile Regression Model for Citation Analysis with Mass Points at Lower Values}



\author[M. Shahmandi, \textit{et al.}]{ Marzieh Shahmandi \orcid{0000-0003-2474-7283}, Paul Wilson \orcid{0000-0002-1265-543X}, \and Mike Thelwall \orcid{0000-0001-6065-205X}}

\affiliation{}{Statistical Cybermetrics Research Group, School of Mathematics and Computer Science, University of Wolverhampton, Wulfruna
Street, Wolverhampton WV1 1LY, UK}




\correspondingauthor{Marzieh Shahmandi}{m.shahmandihounejani@wlv.ac.uk}



\begin{abstract}

Quantile regression presents a complete picture of the effects  on the location, scale, and shape of the dependent variable at all points, not just  the mean. We   focus on two challenges for citation count analysis by quantile regression: discontinuity and  substantial mass points at lower counts. A Bayesian  hurdle  quantile regression model for count data with a substantial mass point at zero was proposed by \cite{King.et.al2019}. It uses quantile regression for modeling the nonzero data and logistic  regression for modeling the probability of zeros versus nonzeros.  We show  that substantial mass points for low  citation counts will nearly certainly also  affect parameter estimation in the quantile regression part of the model, similar  to a  mass point at zero. We update the King and Song model by shifting the hurdle point  past the  main mass points. This  model delivers more accurate quantile regression for moderately to highly  cited articles, especially at quantiles corresponding to values just beyond the mass points, and enables estimates of the extent to which factors influence the chances that an article will be low cited. To illustrate  the potential of this method, it is applied to  simulated citation counts and data from Scopus.

\end{abstract}


\newpage
\section{Introduction}\label{Intro}
Citation analysis can help to estimate the relative importance or impact of articles by counting the number of times that they have been cited by other works. Non-specialists in governments and funding bodies or even researchers in different scientific disciplines sometimes use citation counts to help judge the importance of a piece of scientific research \citep{Meho}. Citation analysis has statistical challenges due to the characteristics of citation counts (a substantial mass point at zero, high right skewness, and heteroskedasticity). Various statistical models have been proposed for citation counts \citep[\eg][]{Seglen,Garanina,Redner,THELWALL2014963,Brzezinski2015,Wan,THELWALL2016622,Eom2011,Shahmandi.et.al2019}, but most have sought to model the conditional mean of citation counts from independent variables.  In other words, they generate a formula for the expected number of citations for given values of research-related parameters, such as article age, topic, and the number of authors.

Quantile regression (QR) is a statistical method proposed by \citet{Koenker.et.al1978} to complement classical linear regression analysis \citep[\eg, ][]{Koenker2001,COAD2008}. Unlike a linear regression where the conditional mean of a dependent variable is modeled, in QR the different conditional quantiles of the dependent variable, such as the median, are modeled based on a set of independent variables. In QR the entire distribution of the dependent variable is related to the set of independent variables. In scientometrics, 
\cite{Danell2011} used QR to investigate whether the future citation rate of an article can be predicted from the author’s publication count and previous citation rate. In the study of nanotechnology publications, QR was used to investigate whether funding acknowledgements influence journal impact factors and citation counts as two dependent variables in two separate models \citep{Wang2015}. \citet{Stegehuis2015} proposed a QR-based model to estimate a probability distribution for the future number of citations of a publication in relation to variables such as the publishing journal's impact factor. \cite{Anauati.et.al2016} assessed the life cycle of articles across fields of economic research through QR. \cite{Ahlgren.et.al2017} used QR to show how some factors, such as the number of cited references, affect the field normalised citation rate across all disciplines. In another study, \citet{Wang2018} used QR models to explore the relationship between SCI (Science Citation Index) editorial board representation and research output of universities (measured by indicators such as the number of articles, total number of citations) in the field of computer science, \cite{Mantyla2019} applied QR at the $0.50$ quantile to indicate how factors such as publication venue and author team past citations influence the number of citations of software engineering papers. \cite{Galiani2019} proposed identifying citation ageing by combining QR with a non-parametric specification to capture citation inflation. Despite this extensive use of QR for citation analysis, the problem of the influence of point masses (low citation counts having high frequencies in a set of articles) has not been fully resolved, undermining the value of the results.\\
The continuity of the dependent variable is important for minimisation of the objective function in QR. A discrete dependent variable leads to non-differentiability of the objective function, resulting in problems deriving the asymptotic distribution of the conditional quantiles. A substantial mass point at zero in the data results in all conditional quantiles less than the proportion of the zeros being equal to zero. In some of the articles cited above, the discontinuity of citation counts was ignored, leading to biased and misspecified estimates for parameters in the model. In other articles, citation counts were normalised by different methods, or a random positive value was added to each citation count to account for the discontinuity. In general, for the case of a discrete variable, jittering proposed by \citet{Machado.et.al2005} is used. In jittering, a random noise in the interval $(0,1)$ is added to each data point to make the data continuous. In the situation of the substantial mass point at zero, researchers frequently focus on the interpretation of the upper quantiles of the dependent variable because the apparent variation in the lower tail might be a consequence of random noise produced by the jittering process. In practice, some important parts of the analysis can be lost. For instance, in the case of the citation counts as a dependent variable, we can lose the information about the effects of factors (as independent variables in the model) on zero or very low cited articles. Therefore, a new methodology related to QR should be considered to tackle these challenges. The approach proposed in this article is an extension of the Bayesian two-part Hurdle QR model of \citet{King.et.al2019}. Having a two-part structure is a fundamental aspect of this model. The two-part model of \citet{King.et.al2019} allows zero and non-zero citations to be modeled separately. The QR part of the model is for modeling the non-zeros, and logistic regression is used for modeling the probability of zeros versus non-zeros. The Bayesian structure of the model assists the estimation of model parameters. In addition, \citet{King.et.al2019} showed another advantage of the application of the Bayesian technique for this model. By simulation, it was shown that the estimates of parameters based on the Bayesian method are more precise in comparison to their classical counterparts even for small sample sizes and when the prior information of the parameters in the model is non-informative. In the case of citation count data, there are frequently substantial mass points at one, two, and three, (and possibly also at greater values) which influence the estimates of parameters in the QR part of the model in a similar manner to the substantial mass point at zero, so a new update of the model will be proposed to reduce the effect of the substantial mass points on the estimation of the model. We take “substantial” to mean greater than about 6$\%$ as mass points less than this appear to have little effect on subsequent estimation (see Table \ref{table:2}). This paper, based on simulations of log-normal continuous data with substantial mass points at zero, one, two, and three (approximating a common distribution of citation counts), will assess, by considering the mean squared error of the estimates of the coefficients corresponding to the independent variables in the model, whether the QR part of the two-part model with a hurdle at three, results in more accurate estimates than are obtained by the other models. We also assess prediction errors and credible intervals for the estimates.

\section{Definitions and concepts}

\subsection{Quantile regression}
\citet{Gilchrist} describes a quantile as \textquotedblleft the value that corresponds to a specified proportion of an (ordered) sample of a population\textquotedblright. The quantiles are the values which divide the distribution such that there is a given proportion of observations below the quantile. Thus the $\tau^{th}$ quantile splits the area under the density curve into two parts: one with area $\tau$ below the $\tau^{th}$ 
quantile and the other with area $1-\tau$ above it. The best-known quantile is the median, which is the $0.50$ quantile. The median is a measure of the central tendency of the distribution: half the data are less than or equal to it and half are greater than or equal to it. In general, for any $\tau$ in the interval $(0,1)$ and any continuous random
variable $Y$ with the probability distribution function $F$, the $\tau^{th}$ quantile of $Y$ can be defined as:

$$F_Y(y_\tau)=P(Y \le y_\tau)=\tau$$

and, the empirical quantile distribution function can be defined as:
$$y_\tau=F^{-1}_Y(\tau)=\inf\{y|F_Y(y)\ge \tau\}.$$

The regression model for the conditional quantile level $\tau$ of $Y$ is \citep{Koenker.et.al1978}

\begin{equation}
Q_{Y_i}{(\tau |\mathbf{x_i}) }=\mathbf{x_i}^{T}\boldsymbol{\beta_\tau} \label{QR}
\end{equation}

where $\mathbf{x_i}$ is $i$th vector of $p$ independent variables, and $\beta_\tau$ is estimated by minimization of the sample objective function or the weighted absolute sum:

\begin{equation}
\displaystyle \min_{\beta_\tau} \sum_{i=1}^{n} \rho_{\tau}\bigg(y_i-\mathbf{x_i}^{T}\boldsymbol{\beta_\tau}\bigg)
\label{SOF}
\end{equation}

where $n$ is the number of observation and $\rho_\tau(r)=\tau \max(r,0)+(1-\tau) \max(-r,0)$ is the check loss function. The solution of Equation (\ref{SOF}) can be obtained by linear programming techniques such as the simplex method \citep{Dantzig1963}, the interior-point method \citep{portnoy.et.al.1997}, and the smoothing method \citep{Clark.et.al.1986,Madsen.et.al.1993}. 
QR preserves $Q_{Y}{(\tau |\mathbf{x})}$ under transformation. Suppose that $\eta(\cdot)$ is a non-decreasing (monotone) function on $\mathbb{R}$, then

$$Q_{\eta(Y)}{(\tau |\mathbf{x}) }= \eta(Q_{Y}{(\tau |\mathbf{x}) })$$

This is important because in the following, transformations need to be used for citation analysis.

\subsection{The Asymmetric Laplace distribution in Bayesian QR} 
Here we define the (three-parameter) Asymmetric Laplace Distribution. Because of the distribution-free characteristic of QR, the minimisation of Equation (\ref{SOF}) can be considered as a non-parametric problem. This can cause a challenge for defining the Bayesian version of QR because the Bayesian framework needs the likelihood function of the model. Different approaches have been suggested for this issue, but the ALD method proposed by \citet{Yu.et.al.2001} is the simplest and most understandable method. The ALD has density probability function:

\begin{equation}
f{(y_i |\mu,\sigma,\tau)}= \frac{\tau(1-\tau)}{\sigma} ~\mbox{exp}{\Big \{ \rho_{\tau}\bigg(\frac{y_i-\mu}{\sigma}\bigg)\Big \}}
\label{Laplace}
\end{equation}

where $\mu \in \mathbb{R}, ~\sigma>0$, and $\tau \in [0,1] $ are  respectively location, scale, and skewness parameters.

For a random variable $W$ where $W \sim  \mathcal{ALD}(\mu,\sigma,\tau)$, there is a location-scale mixture representation following a normal distribution with specific parameters \citep[\eg, ][]{Lee.et.al2010,Kozumi.et.al2011}. In fact:

\begin{equation}
W_i |v_i \sim \mathcal{N}(\mu+ \theta v_i,\psi^2\sigma v_i)
 \label{QR_N_nonzero}
\end{equation}

where 
$$\theta=\frac{1-2\tau}{\tau(1-\tau)}, ~~~\psi^2=\frac{2}{\tau(1-\tau)}.$$

Two variables $u$ and $v$ are independent. $u$ follows a standard normal distribution, and $v$ is exponentially distributed with mean $\sigma$. This valuable feature of ALD enables the use of QR in the Bayesian framework.

By considering the location parameter in ALD as a linear function of the independent variables, $\mu_i=\mathbf{x_i}^{T}\boldsymbol{\beta_\tau}$, 
the maximum likelihood estimate of the $\beta$ in Equation (\ref{Laplace}) is equivalent to the estimate obtained from the minimisation of Equation (\ref{SOF}), for every fixed $\tau$. QR may be regarded as linear regression where the error term has been replaced by the ALD distribution. ALD provides a likelihood base for data in the Bayesian framework which holds the data fixed, and treats the parameters as random variables which are explained probabilistically by prior knowledge. The combination of the evidence extracted from the data (likelihood) and the prior beliefs is a posterior distribution corresponding to the parameters. A Gibbs sampler of the Markov chain Monte Carlo (MCMC) method is used for the approximation of the posterior distribution. 

\section{Bayesian two-part hurdle QR}
\citet{Santos.et.al2015} proposed a Bayesian two-part QR methodology for a continuous response variable with a substantial mass point at zero or one. \citet{King.et.al2019} introduced a Bayesian two-part QR model with a hurdle at zero for the case of count data with a substantial mass point at zero. In the following, a new version of this model for the case of a hurdle at a specific value of $c$ is introduced.  To fit this model, for the first step, the count data should be transformed by

\begin{equation}
y^{*}_i = \left\{
    \begin{array}{ll}
0 &\quad ~y_i\leq c \\
\ln{(y_i-c-u_i)} & \quad \;\; y_i>c
    \end{array}
\right.
 \label{tranformation}
\end{equation}

to provide a semi-continuous variable. By this transformation, all substantial mass points less than or equal to $c$ are mapped to zero and the rest of the data are converted to a real number in the domain of the ALD distribution. By considering $c=0$ in the relationship (\ref{tranformation}), the original transformation used by \citet{King.et.al2019} is obtained. 


The two-part probability function has the form 

\begin{equation}
f(y^*_i|\boldsymbol{\gamma},\boldsymbol{\beta_\tau}, \sigma,v_i, \tau)=(\omega_i)~.~\mathbb{I}(y^*_i=0)+(1-\omega_i).~\mathcal{N}(\mathbf{x_i}^{T}\boldsymbol{\beta_\tau}+ \theta v_i,\psi^2\sigma v_i)~\mathbb{I}(y^*_i \neq 0)
 \label{two-part}
\end{equation}

where $\omega_i=P(y^*_i = 0)$ and $\mathbb{I}$ is the indicator function. In the literature the parameter $\omega$ has been used both to denote the probability of observing a non-zero (for example  \citet{King.et.al2019}), and a zero, (for example \citet{Santos.et.al2015}, \citet{Ospina.al.2012}).  Previous research in the area of citation count analysis, for example \citet{Didegah.al.2013}, has used $\omega$ to denote the probability of observing a zero, thus we shall follow this precedent here.

A logit link is usually applied to model $\omega_i$ based on a linear combination of the independent variables so that

\begin{equation}
\mbox{logit}(\omega_i)=\mathbf{z_i}^{T}\boldsymbol{\gamma}
\label{logit_tp}
\end{equation}

where $\mathbf{z_i}$ is a vector of independent variables. The variables used to model $\omega_i$ may or may not be the same as those used to model the non-zero data.

The two-part model is a mixture model that is a linear combination of a continuous normal distribution (corresponding to QR for modelling the jittered non-zero citation counts) and a point distribution at zero.  $\omega_i$ and $1-\omega_i$ are respectively the contributions of the point distribution and the continuous distribution in this mixture. This is a hurdle model because the zeros and non-zeros are modeled separately.

From Equation (\ref{logit_tp}), Equation (\ref{two-part})
can be rewritten as 

\begin{align}
f(y^*_i|.) &=\omega_i^{\mathbb{I}(y^*_i=0)} \Big [(1-\omega_i).~\mathcal{N}(\mathbf{x_i}^{T}\boldsymbol{\beta_\tau}+ \theta v_i,\psi^2\sigma v_i)\Big ]^{\mathbb{I}(y^*_i \neq 0)} \nonumber \\[4mm]
      &= \omega_i^{\mathbb{I}(y^*_i=0)} (1-\omega_i)^{\mathbb{I}(y^*_i \neq 0)}\Big [ \mathcal{N}(\mathbf{x_i}^{T}\boldsymbol{\beta_\tau}+ \theta v_i,\psi^2\sigma v_i)\Big ]^{\mathbb{I}(y^*_i \neq 0)} \nonumber \\[4mm]
       &= \bigg[\frac{1}{1+\mbox{exp}(-\mathbf{z_i}^{T}\boldsymbol{\gamma})}\bigg]^{\mathbb{I}(y^*_i=0)}\bigg[\frac{1}{1+\mbox{exp}(\mathbf{z_i}^{T}\boldsymbol{\gamma})}\bigg]^{\mathbb{I}(y^*_i\neq 0)}  \label{likeli} \\
       &~~~~~~\times \Big [\mathcal{N}(\mathbf{x_i}^{T}\boldsymbol{\beta_\tau}+ \theta v_i,\psi^2\sigma v_i)\Big ]^{\mathbb{I}(y^*_i \neq 0)} \nonumber
       
\end{align}

Suppose non-informative priors:\\
$\pi({\boldsymbol{\beta_\tau}}) \sim \mathcal{N}(\tilde{b},\tilde{B})$\\
$\pi({v_i}) \sim \mathcal{E}(\sigma)$\\
$\pi({\sigma}) \sim \mathcal{IG}(\tilde{n},\tilde{s})$\\
$\pi({\boldsymbol{\gamma}})  \sim \mathcal{N}(\tilde{g},\tilde{G})$\\

where $\mathcal{E}$ denotes the exponential distribution with mean $\sigma$ and $\mathcal{IG}$ denotes an inverse gamma distribution with the hyperparameters $\tilde{n}$ and $\tilde{s}$. The posterior distribution of the model is

\begin{equation}
\pi(\boldsymbol{\beta_\tau},\boldsymbol{\gamma},\sigma,v_i |y^*)\propto L(y^*_i|\boldsymbol{\beta_\tau},\sigma,v_i,\tau)\pi(\boldsymbol{\beta_\tau}) \pi(\boldsymbol{\gamma}) \pi(\sigma) \pi(v_i|\sigma)
 \label{BQR_Frame_m}
\end{equation}

where $L(y^*_i|.)$ is the likelihood function of $f(y^*_i|.)$. To approximate the posterior distribution, the Gibbs sampler of the MCMC method will be used.

\section{Simulation study}
\begin{figure}[H]
  \centering
   \includegraphics[width=.9\linewidth]{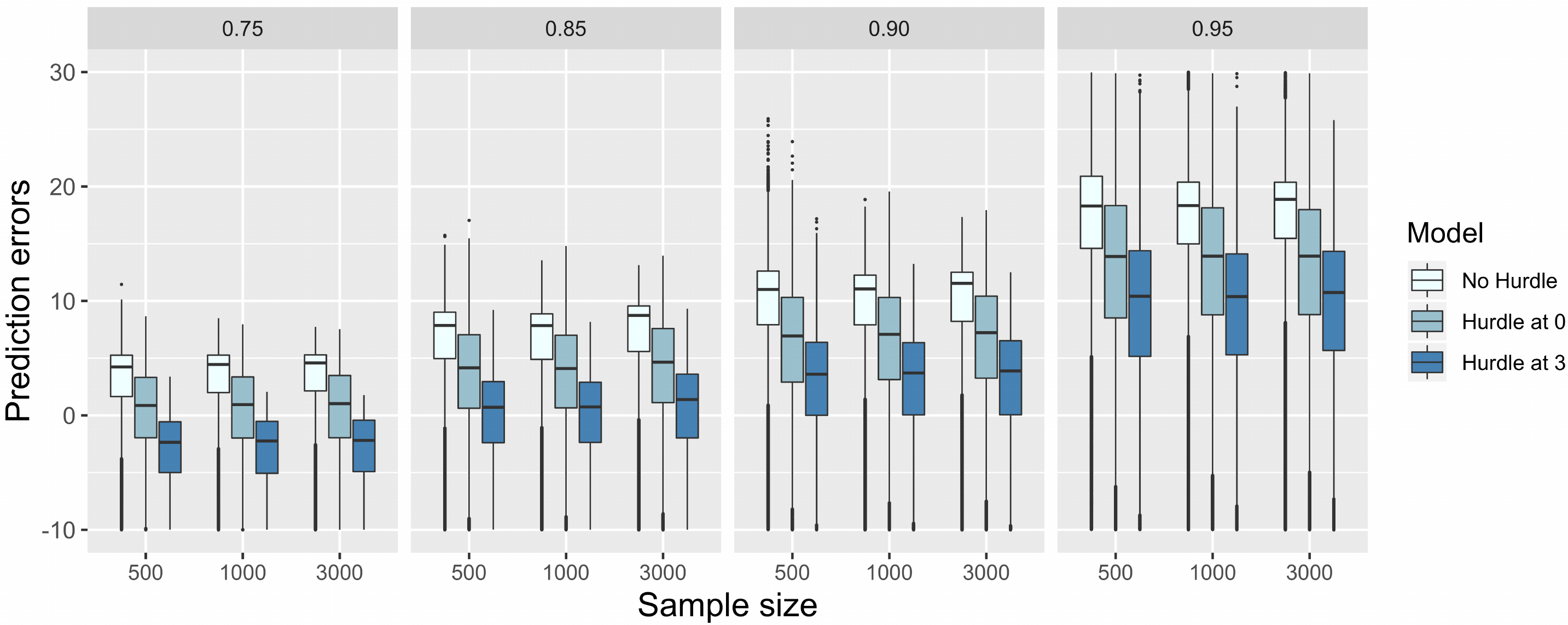}
  \caption{Prediction errors based on different models and sample sizes}

  \label{fig1}
\end{figure}

\begin{figure}[H]
  \centering
   \includegraphics[width=.9\linewidth]{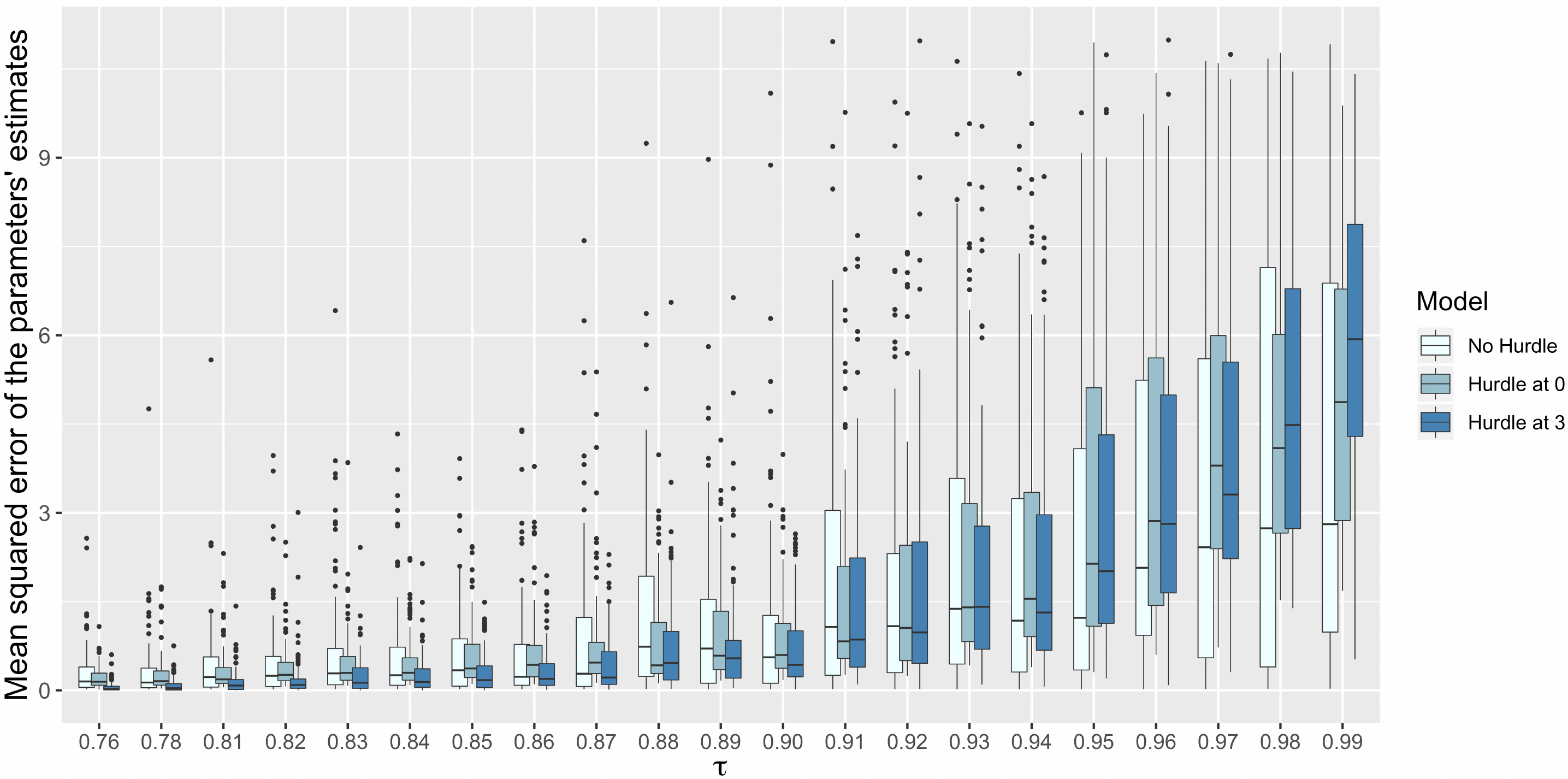}
  \caption{Mean squared errors for the parameter estimates (excluding intercept) for different models with a sample size of 1000}

  \label{fig1_1}
\end{figure}

\begin{figure}[H]
  \centering
   \includegraphics[width=.9\linewidth]{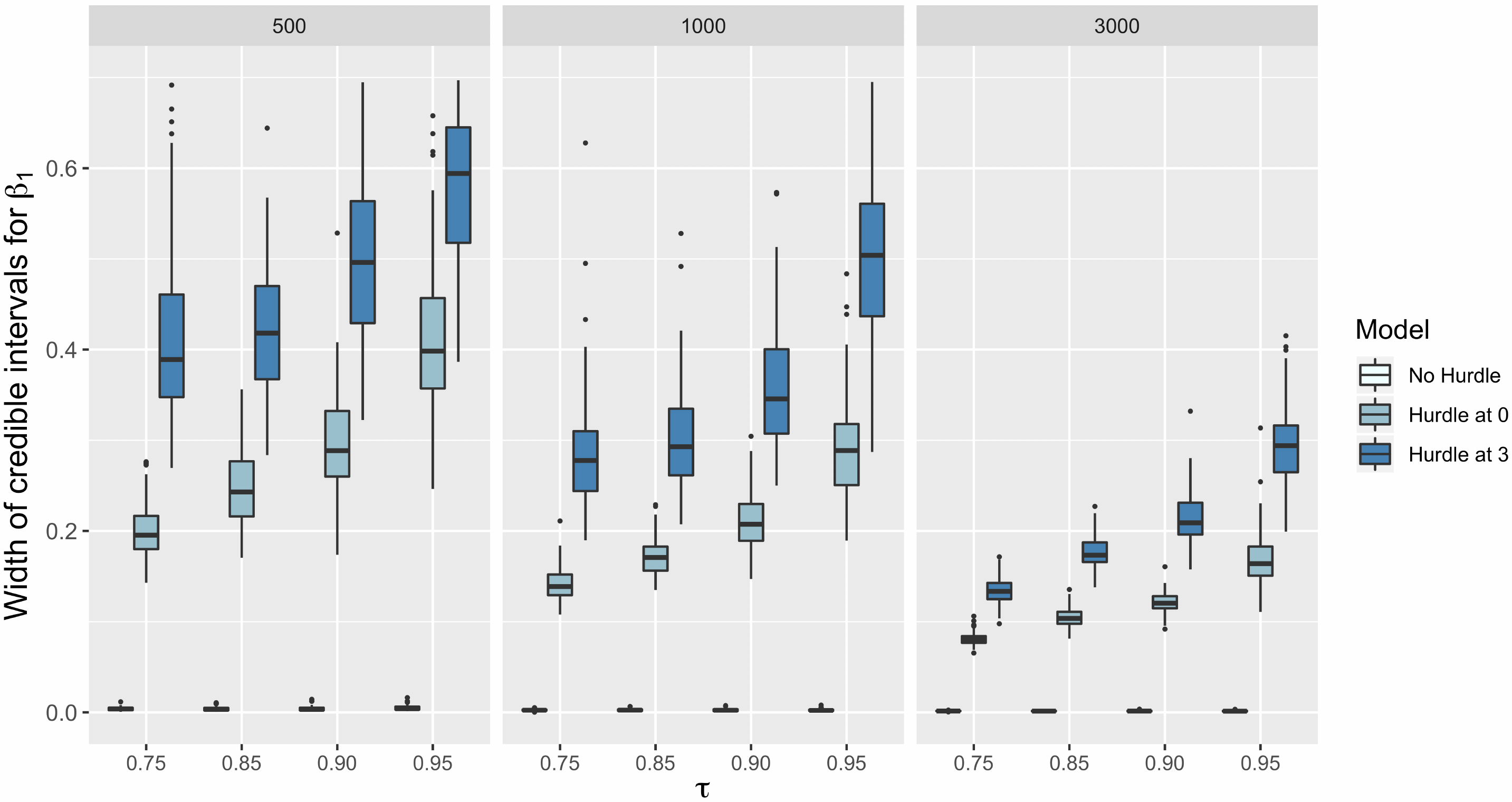}
  \caption{Width of credible intervals for the estimates of $x_1$}

  \label{fig2}
\end{figure}

\begin{figure}[H]
  \centering
   \includegraphics[width=.9\linewidth]{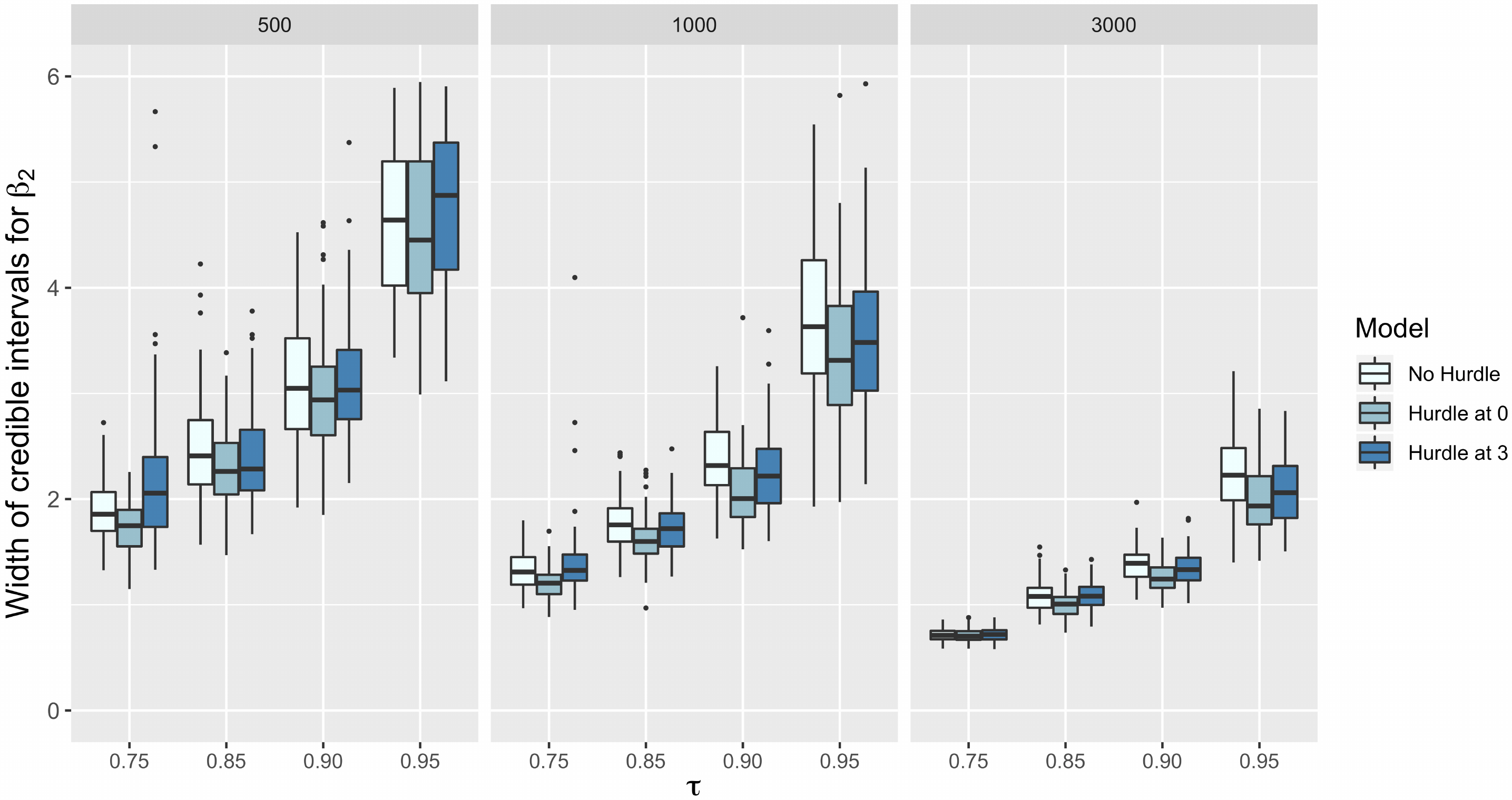}
  \caption{Width of credible intervals for the estimates of $x_2$}

  \label{fig3}
\end{figure}

In this section, samples with sizes of 500, 1000 and 3000 are simulated from continuous log-normal distribution ($\mathcal{LN}$) with mean $(2 -0.2*x_1+0*x_2 +\epsilon)$ and standard deviation $0.4$ where $x_1 \sim \mathcal{LN}(2,2)$ , $x_2 \sim \mathcal{N}(0.5,0.5)$, and $\epsilon \sim \mathcal{N}(0,1)$. The log-normal distribution was chosen because it approximates the typical distribution of citation counts. The floor function was used for the simulated values less than 4 to simulate substantial mass points at $0$, $1$, $2$ and $3$. The intercept value of $2$ and the coefficient $-0.2$ of $x_1$ were chosen so that approximately 45\% of the data will be zeros and 75\% of the data less than 3, similar to much citation count data. The coefficient of $x_2$ was chosen as zero to enable comparison of the proposed models when one of the variables is non-significant. Bayesian QR and Bayesian two-part QR models with hurdle at 0, and with hurdle at 3 will be fitted. The objective is to compare Bayesian QR with the QR parts of the two-part models with hurdles at 0 and 3. For each sample size and for each quantile level, the Bayesian QR model is fitted to the whole data. Then the quantile level of the corresponding quantile value is found in the data in which the zeros are excluded and the Bayesian QR model is fitted, (i.e. the QR part of the two-part model with hurdle at 0). Next, the quantile value corresponding to the quantile level is found in the data in which all substantial mass points (including 0,1,2, and 3) are removed and the Bayesian QR model is fitted for the corresponding quantile (this model is the QR part of the two-part model with hurdle at 3). For more clarification, suppose the specific quantile level is $0.85$, and the Bayesian QR model is fitted to the whole data at this quantile. Say the value corresponding to this quantile is $7.640$. Now the quantile corresponding to $7.640$ for the data with zeros removed is computed; say it is $0.70$. Then the Bayesian QR model is fitted to this data ($>0$) for the $0.70$ quantile. This model is the QR part of the two-part model with a hurdle at 0. Next, the quantile level corresponding to $7.640$ is found in the data with substantial mass points at 0,1,2, and 3 removed. Say this is $0.40$. The Bayesian QR model is fitted to this data ($\geq3$) at this quantile. The estimates of parameters based on this model are the QR part of the two-part model with a hurdle at 3. This process is repeated 100 times for each sample size, and for 4 specific quantiles of $0.75$, $0.85$, $0.90$, and $0.95$ (of the whole data) separately. For each combination, the prediction error of each model, the mean squared error of the parameters' estimates (intercept excluded), and the width of the credible intervals for both independent variables $x_1$ and $x_2$ are computed. The function $jags$ from the \textsf{R}-library $R2jags$ which is based on Gibbs sampler was used for MCMC computation corresponding to the Bayesian QR models. In this function, three chains will run. For each chain, 10000 iterations with burn-in 1000 and thinning number of 90 were considered. The quality of the obtained MCMC samples was assessed based on both qualitative (graphical) and quantitative diagnostics. For example, autocorrelation plots showed by increasing the lag number, the correlation between the samples decreases sharply and approaches zero, indicating the independence among the samples. In addition, the Gelman-Rubin potential scale-reduction factor (PSRF) diagnostic and its values near 1 showed achieving convergence in the MCMC chains. The effective sample size diagnostic revealed the reasonable number of independent samples for parameters of each model.  \textsf{R} code related to the simulations are available online \footnote{https://doi.org/10.6084/m9.figshare.13726198.v1}.
Finally, for each model, the boxplots of the prediction errors, the mean squared errors of the parameters' estimates, and the width of the credible intervals are compared. The results are reported in Figure\, \ref{fig1}, Figure\, \ref{fig1_1}, Figure\, \ref{fig2}, and Figure\, \ref{fig3}.

Figure\, \ref{fig1} shows the prediction errors $(y-\hat{y})$ where $\hat{y}$ is calculated for each fitted model for each quantile and sample size. It shows that the prediction error for the QR part of the two-part model with hurdle at 3 is least, followed by the model with hurdle at 0, and then followed by the Bayesian QR (no hurdle model).



Figure\, \ref{fig1_1} displays the mean squared errors of the parameters' estimates computed based on the formula:
$$MSE=\frac{1}{p}\sum_{i=1}^{p} (\beta_i-\hat{\beta_i})^2$$

where $p$ is the number of independent variables, not including the intercept, in the model. All calculations were based on a sample size of 1000 and quantile levels of $0.76$, $0.78$, and a sequence from $0.81$ to $0.99$ (of the whole data). The first quantile was chosen as $0.76$ because its corresponding value is just greater than three. This set of quantile levels are considered to present a complete picture of the trend of the mean squared errors corresponding to all three models. Smaller values (near zero) of the mean squared errors are desirable. The results show that, in general, for quantile levels lower than $0.93$, the hurdle model at 3 has outperformed the hurdle model at 0. In addition, the hurdle models present more precise estimates in comparison to the no hurdle model. However, for the cases of quantile levels greater than $0.93$, there are examples that the hurdle model at 0 has the poorest estimates in comparison to the other two models. Just for one quantile level ($0.99$), the hurdle model at 3 shows the bigger mean squared errors. It can be deduced that 
influence of the mass points is greatest at the quantiles shortly after the mass points (where the hurdle models particularly the hurdle model at 3 show more accurate estimates), but by the time we reach the extreme upper quantiles the influence has waned and the no hurdle model returns better estimates as it is based upon a larger sample size. In general, by increasing the quantile, the estimates of the mean squared errors become larger for all three models.

\newpage
Figure\, \ref{fig2} illustrates the width of credible intervals for the estimates of the coefficients of $x_1$ based on the different models and sample sizes. The credible intervals provided are based on the percentiles of the posterior probability distribution. Credible intervals are the Bayesian counterparts of confidence intervals in classical statistics. We see that, as is to be expected, by increasing the sample size from 500 to 3000, the width of the credible interval decreases considerably. Moreover, the model with hurdle at 3 has the largest width for all the quantiles, followed by the model with hurdle at 0, followed with the no hurdle model. Again, this is as would be expected as less data is available for the hurdle at 3 model than for the hurdle at 0 than for the no hurdle model.



Figure\, \ref{fig3} shows the widths of the credible intervals for the estimates of the coefficients of $x_2$ based on the different models and sample sizes. The coefficient of $x_2$ in the model from which the data was simulated was $0$. That is, $x_2$ is not significant. The figure shows that for most of the quantiles and sample sizes, the widths of credible intervals based on the three models are approximately the same. This approximate equality of widths of credible intervals across the models is surprising given that the models all have different amounts of data available to them. It is unclear whether this phenomenon is universal, or whether it only applies to specific data, this approximate equality of the widths of credible intervals has also occurred in other simulations performed by the authors however, and is worthy of further investigation.

The results of the simulation show that the model with hurdle at 3 in general returns more accurate estimates based on the mean squared errors of the estimates of parameters (excluding the intercept) at quantiles just beyond the hurdle. The hurdle models mostly can present the more precise estimates for the parameters for moderate and high levels of quantiles while for the extreme high quantiles, the no hurdle model has better estimates due to waning the biasing influence of the mass points by this point, and the fact that the no-hurdle model works off a greater amount of data. In addition, the model with hurdle at 3 results in smaller prediction errors at the cost of wider credible intervals. This followed with the model with hurdle at 0, then followed with the no hurdle model. Moreover, a larger sample size also decreases the differences between models for the width of their credible intervals (particularly when the independent variable is not significant in the model).

\newpage

\section{Citation count example} \label{citation_lab}
The data used in this article consists of citation counts for standard journal articles (excluding reviews) published in the following seven Scopus fields: \textsl{Arts and Humanities (all)}, \textsl{Literature and Literary Theory}, \textsl{Religious Studies}, \textsl{Visual Arts and Performing Arts}, \textsl{Media Technology}, \textsl{Architecture}, and \textsl{Emergency Nursing}. The articles were published in the year 2010 and their information was extracted at the end of the year $2019$, giving the citation counts time to mature. These seven fields were selected because after discarding the records with missing cells for computation of the dependent and possible independent variables, they have the highest proportions of zeros and also they have maximum records of 3000. The effect of sample size on computation time is a method limitation because MCMC is time-consuming. The number of citations to each article is the dependent variable. The independent variables available were the number of keywords, the number of pages, title length, abstract length, collaboration (the number of authors of an article), international collaboration, abstract readability and journal internationality. Collaboration, length of title, and journal internationality were selected as independent variables because with this selection fewer records with missing data had to be discarded and the percentage of zeros in the data remained high. The selected variables have a reasonably strong correlation with the corresponding citation counts for most of the seven fields. \\

\begin{center}
\begin{table}[hhb]{{\extracolsep{\fill}}{}}
\caption{\label{table:2}  Details of the citation count data for the seven fields from 2010 analysed}
\scriptsize{\resizebox{1\textwidth}{!}{\begin{tabu}{|c|c|c|c|c|c|c|c|}
\hline

\multirow{3}{*}{Field}~~~& ~~~\multirow{3}{*}{Number of articles}~~~&~~\multirow{3}{*}{Percentage of zeros}~~~&~~\multirow{3}{*}{Percentage of ones}~~&~~\multirow{3}{*}{Percentage of twos}~~&\multirow{3}{*}{Percentage of threes}~~&~~Percentage of &~~Total percentage of\\ 

~~~& ~~~~~~&~~~~~&~~~~&~~~~&~~&~~substantial mass points&~~substantial mass points\\ 

~~~& ~~~~~~&~~~~~&~~~~&~~~~&~~&~~in $1  \leq y \leq 3$ &~~\\ \hline

Literature and Literary Theory& 3126& 51&18 &11 &5&34 & 85\\
Arts and Humanities&1460 &41 &16 & 8&7& 31 &72\\
Visual Arts and Performing Arts& 1799&39 &16 & 10&6&32  &71\\
Architecture&2215 &36 &16 &10& 7&33 &69\\
Religious Studies&2176 &32 & 18& 11&8& 37 &69\\
Emergency Nursing& 1299& 39& 10& 6& 5& 21&60\\
Media Technology&1889 & 27& 8& 6&6& 20 &47\\

\hline
\end{tabu}}}
\end{table} 
\end{center}

The highest proportions of zeros are related to both \textsl{Literature and Literary Theory} and \textsl{Arts and Humanities} respectively (Table \ref{table:2}). The portions of ones, twos, and threes in comparison to the portion of zeros are not huge but they are still noticeable. There are no substantial mass points greater than 3 for the fields. The percentage of substantial mass points greater than zero in the fields varies from approximately 20$\%$ for \textsl{Media Technology} up to 37$\%$ for \textsl{Religious Studies}. The total percentages of substantial mass points for the fields of \textsl{Literature and Literary Theory}  and \textsl{Media Technology} are the largest (85$\%$) and smallest (47$\%$) respectively. \\
Collaboration, title length, and journal internationality were included in the models as independent variables. Collaboration and title length are discrete variables. The log function of collaboration was used in the models to provide a closer to linear relationship with the citation counts. Journal internationality is a continuous variable on the interval $[0,1]$. Journal internationality was computed with the Gini coefficient \citep{Corrado.1997}. A value of $0$ shows the highest level of internationality of the journal related to the article, and a value of $1$ shows the least internationality. 
A sequence of quantiles from $0.05$ to $0.95$ is considered. Ordinary Bayesian QR, Bayesian two-part QR with hurdle at 0 and Bayesian two-part QR with hurdle at 3 were fitted to the datasets. MCMC was calculated again with the $jags$ function by considering three chains. For each chain, 100000 iterations with burn-in 50000 and thinning size of 160 were used. Based on a pre-test, the autocorrelation plots for the parameters corresponding to the journal internationality in both parts of the model showed a slow decreasing pattern, indicating slow mixing in the chain. To fix it, such a large number of iterations, burn-in and thinning size were selected. The qualitative (graphical) and quantitative convergence diagnostics the same as the ones used for simulation section, were applied for checking the quality of the MCMC samples. The autocorrelation plots decreased and approached zero by increasing the lag number, showing convergence in the chains. There are also reasonable values for the effective sample size and also PSRFs near 1 for all parameters in the models. \textsf{R} code related to the MCMC computation is available online \footnote{https://doi.org/10.6084/m9.figshare.14742939.v4}.\\

\textbf{Comparison of the QR part of the two-part models with the Bayesian QR model}\\
In the following, the results related to the QR parts of the Bayesian two-part QR models with hurdle at 0, and 3 are compared to the results of the Bayesian QR.\\

\begin{figure}[H]
  \centering
   \includegraphics[width=.9\linewidth]{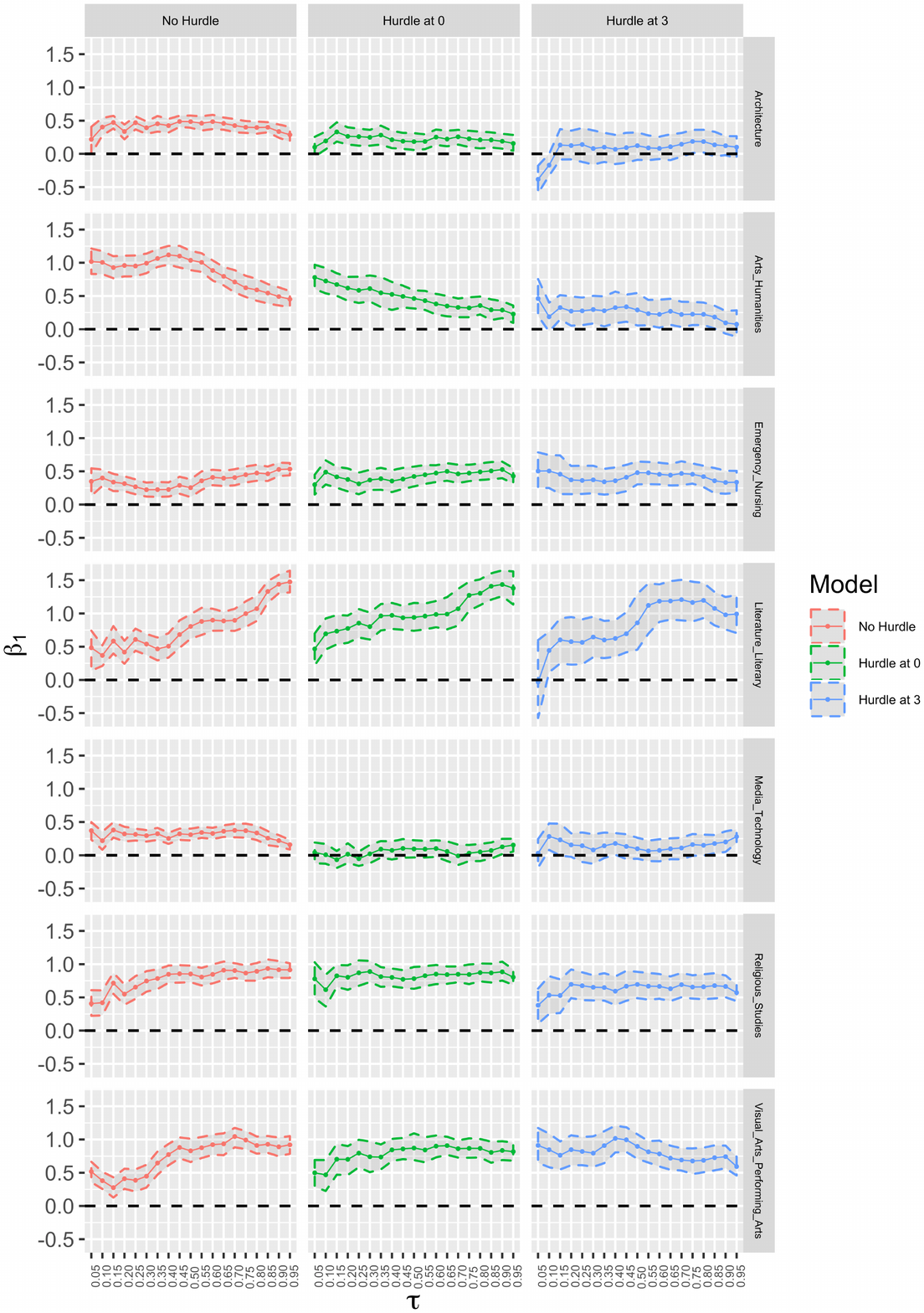}
  \caption{The parameter estimates for collaboration ($\beta_1$) over the quantiles of the citation count distribution separated by types of the models
 }
 
 
  \label{fig5}
\end{figure}

In Figure\,\ref{fig5}, the linear effect of collaboration and its $95\%$ credible intervals over all the quantiles of the citation counts in different fields are shown. The credible intervals provided based on the percentiles of the posterior probability distribution in Bayesian statistics are counterparts of confidence intervals in classical statistics. The upper and lower boundaries of the $95\%$ credible intervals are represented by dashed lines. A narrower band illustrates a smaller variance for the estimated parameter. When a band includes zero, it indicates a non-significant effect related to the variable.

\newpage
\newpage

The effects of collaboration on citation counts in Bayesian QR are significantly positive at all quantiles for all fields. In comparison with two-part models with hurdle at 0 and 3, in the Bayesian QR model, the impact size of the collaboration for \textsl{Literature and Literary Theory} and \textsl{Emergency Nursing} was the smallest over the quantiles while for \textsl{Architecture}, \textsl{Media Technology}, \textsl{Religious Studies}, and  \textsl{Arts and Humanities}, the effect size is the largest. Particularly for the field of \textsl{Arts and Humanities} that has substantial mass points at lower counts, the difference in impact size based on the Bayesian QR model against the hurdle models is considerable. For the field of \textsl{Visual Arts and Performing Arts} for the first half of the quantiles, the effect size of collaboration for the Bayesian QR model is the smallest, but for the second half of the quantiles, it is the greatest. By discarding the substantial mass points of the citation counts and fitting the models with hurdle at 0 and 3, the collaboration effect stabilises over the quantiles for most of the fields (except \textsl{Literature and Literary Theory} and \textsl{Arts and Humanities}), indicating collaboration equally influences the moderately-cited and highly-cited articles. For \textsl{Literature and Literary Theory}, the effect still follows an increasing trend based on the hurdle models the same as the no hurdle model, showing the most benefit of collaboration for the highly-cited articles. However, for \textsl{Arts and Humanities}, based on all three models, collaboration experiences a downward trend.
In previous studies, collaboration has sometimes (but not always) been shown to be related to citation counts. Previous research has used different data sets and statistical methods to assess the relationship between citation counts and collaboration with differing results. For example, \citet{Bornmann_2012} used a negative binomial regression model (with a log link) for approximately 2000 manuscripts that were submitted to the journal Angewandte Chemie International Edition (AC-IE). The estimated coefficient of collaboration was $0.023$ (i.e. for each unit increase in collaboration, the log of the citation count increases by $0.023$ on average) with a $p$-value greater than $0.05$. However, \citet{borsuk2009a} used ordinary least squares (OLS) regression to analyse data of six journals in ecology from 1997 to 2004 and estimated that the effect size and $p$-value were $0.196$ and $0.005$ respectively. By applying the negative binomial hurdle model, \citet{Didegah.2014} also showed that the collaboration has a significant positive impact on citation counts for all subjects of the Web of Science except \textsl{Physics}.

\begin{figure}[H]
  \centering
   \includegraphics[width=.9\linewidth]{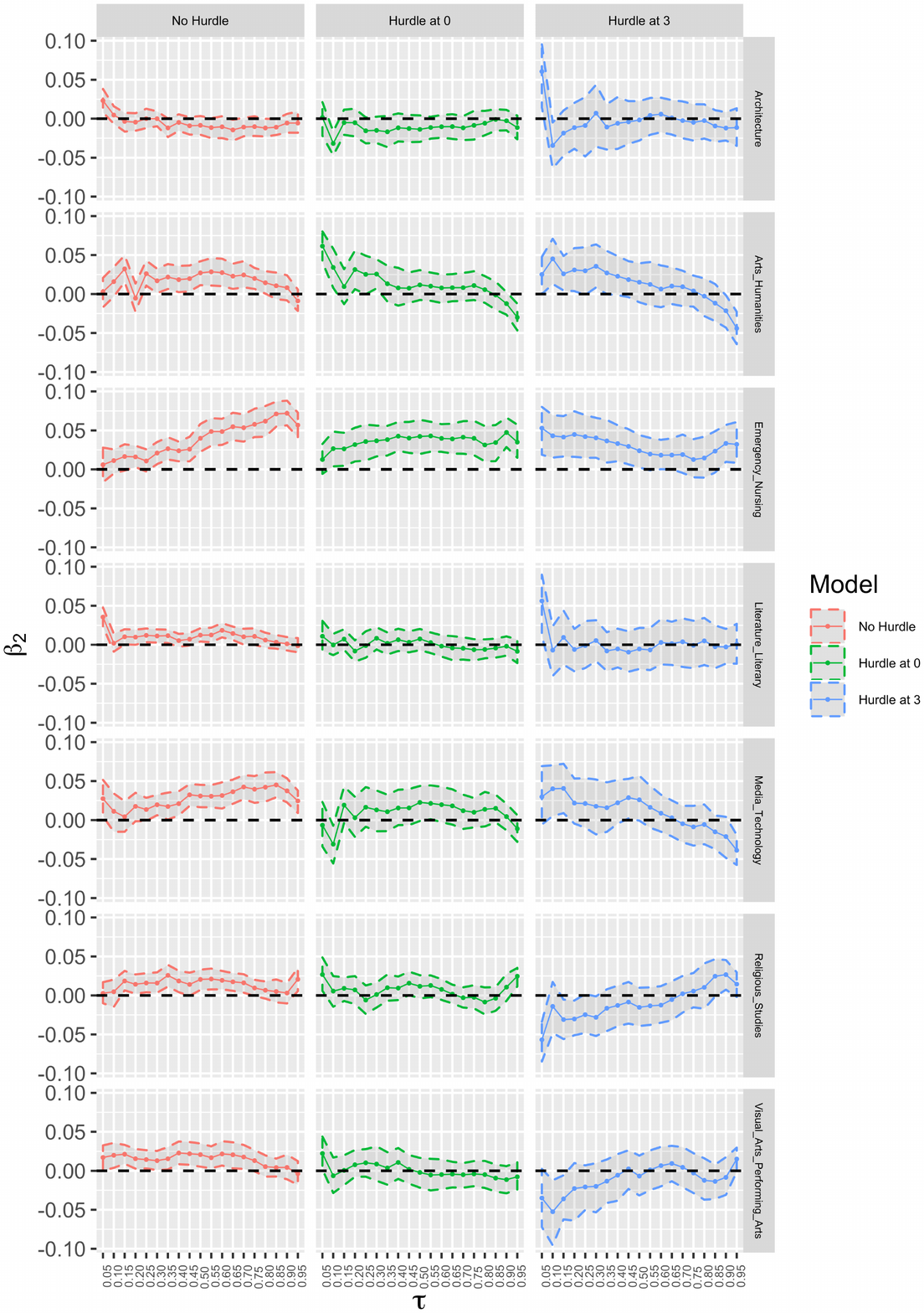}
  \caption{The parameter estimates for title length ($\beta_2$) over the quantiles of the citation count distribution separated by types of the models
 }
 
  \label{fig7}
\end{figure}

Figure\, \ref{fig7} displays the linear impact of the length of title on the citation count distribution across the quantiles in the various fields. Based on the Bayesian QR model, in general, the effect is mostly positive but really small in size and just statistically significant for some quantiles in some fields. According to this model, this effect fluctuated gradually over the quantiles, illustrating that low, moderately, and highly cited articles are equally influenced by this effect. The impact size of the length of title based on the Bayesian QR is greatest for most of the fields and quantile levels in comparison to the two-part QR models with hurdle at 0, and hurdle at 3. This effect is a significant factor for just a few numbers of fields and quantiles based on the two-part models with hurdle points of 0 and 3. By skipping the zero mass point and fitting the Bayesian two-part QR model with hurdle at 0, the effect shows a flatter pattern in comparison to the Bayesian QR model particularly for the fields of \textsl{Emergency Nursing} and \textsl{Media Technology}. The effect based on the two-part model with hurdle at 3 shows a slightly different pattern but still with small size for some quantiles in some fields. For example, it shows the least effect size for lower quantiles of the citation counts for the fields of \textsl{Visual Arts and Performing Arts} and \textsl{Religious Studies} that have high percentages of mass points in $1\leq y \leq 3$. When interpreting the various diagrams of Figure \ref{fig7} comments made elsewhere in this paper concerning the relative suitablity of the various models at the various quantiles should be considered, different models being more suitable depending upon the quantile under consideration. According to the previous study of \citet{Haslamet.al.2008} and using correlations tests and regression analyses, longer title lengths displayed a negative impact on citation counts in psychology. In addition, by applying negative binomial hurdle models in different subjects of Web of Science, \citet{Didegah.2014} showed that the mean length of title associated negatively to nonzero citation counts in some fields of Web of Science such as \textsl{Economics $\&$ Business}, \textsl{Computer Science}, and  \textsl{Chemistry}, but non-significantly in the fields of \textsl{Clinical Medicine}, \textsl{Multidisciplinary} and \textsl{Physics}.

\begin{figure}[H]
  \centering
   \includegraphics[width=.9\linewidth]{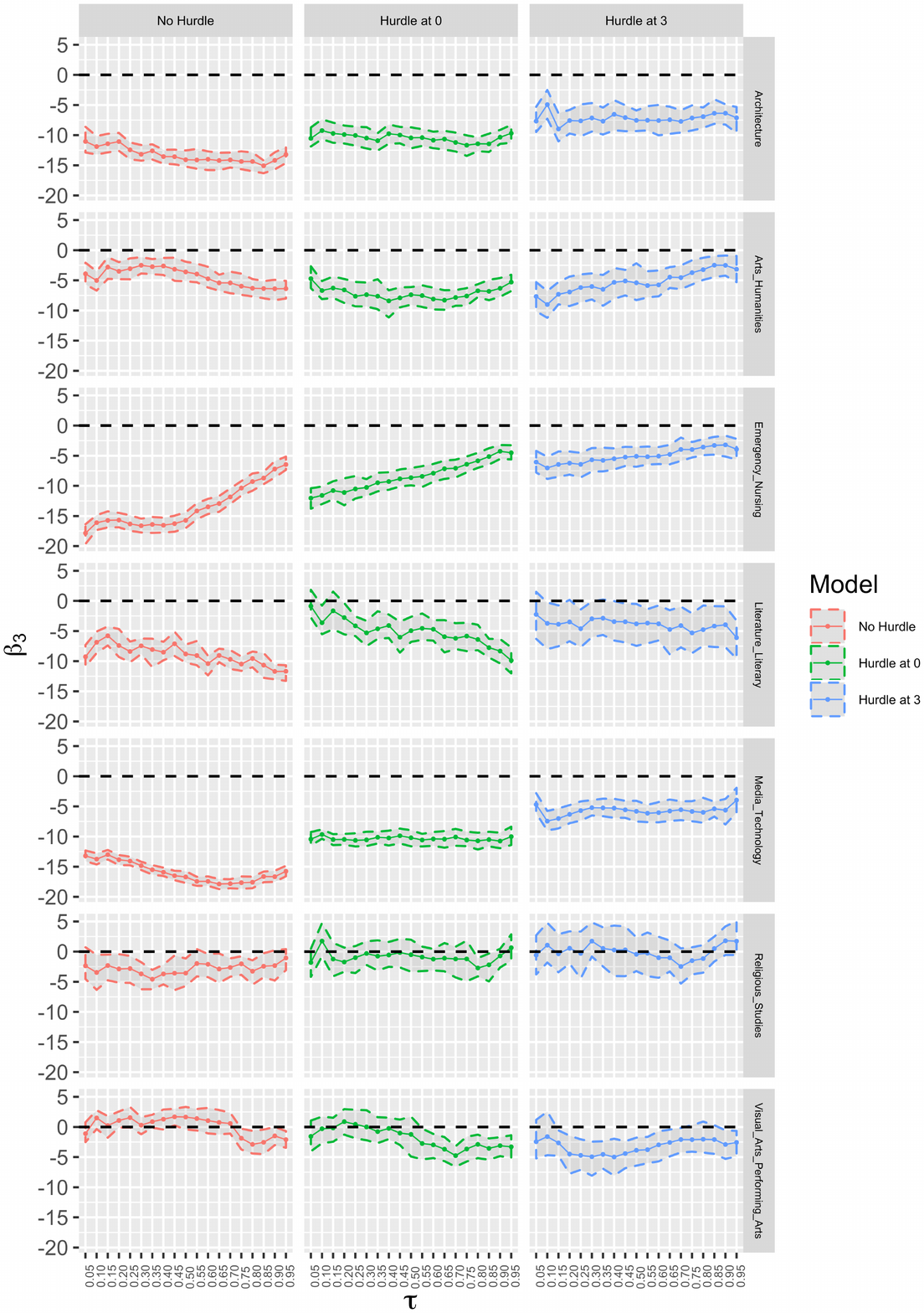}
  \caption{The parameter estimates for journal internationality ($\beta_3$) over the quantiles of the citation count distribution separated by types of the models
 }
 
  \label{fig9}
\end{figure}

\newpage
Figure\,\ref{fig9} illustrates how journal internationality influences the citation counts at all quantiles in the different fields. As was mentioned, a lesser value for the Gini coefficient corresponds to greater journal internationality, indicating that the journals in this field published articles from a broad range of countries. Based on the Bayesian QR, the effect of Gini coefficient significantly negatively influences the citation counts over all the quantiles for all the fields except  \textsl{Visual Arts and Performing Arts} where its impact is not significant for the majority of the quantiles. The negativity of the effect reflects the direct relationship between journal internationality and citation counts.
Fitting the Bayesian two-part QR models with hurdle at 0 and hurdle at 3 results in the trend of the effect becoming smoother and of noticeably smaller magnitude, especially for the model with hurdle at 3, for the quantiles in all fields except \textsl{Visual Arts and Performing Arts} and \textsl{Arts and Humanities} where the estimates based on the various models intersect. Perhaps it refers to the existence of the high portions of mass points in these two fields that influenced the estimates of the effect in the Bayesian QR model. In fact, for the case of \textsl{Visual Arts and Performing Arts}, the Bayesian QR model shows that journal internationality is not significant at most quantiles, whereas the model with hurdle at 3 indicates that it is, the model with hurdle at 0 being somewhere in between. This is a good example that shows the importance of using the appropriate model. Based on the Bayesian QR model, the effect follows mostly a decreasing trend by increasing the quantiles for most of the fields, indicating higher impact size on highly cited articles, but mostly a stabilised trend for the two-part models, showing the equal importance of the effect on moderately and highly cited articles. Previous literature has also found a significant positive
association between journal internationality and citation impact with application of Structural Equation Modelling and simple correlation coefficient by \citet{Yue.2004} and \citet{Kim.2010} respectively. \citet{Didegah.2014} applied the negative binomial hurdle model to show both negative and positive relationships between journal internationality and non-zero citation counts in different fields. For example, positive association in \textsl{Psychiatry/Psychology} subject but negative one in \textsl{Social Sciences} were reported.

\begin{figure}[H]
  \centering
   \includegraphics[width=.9\linewidth]{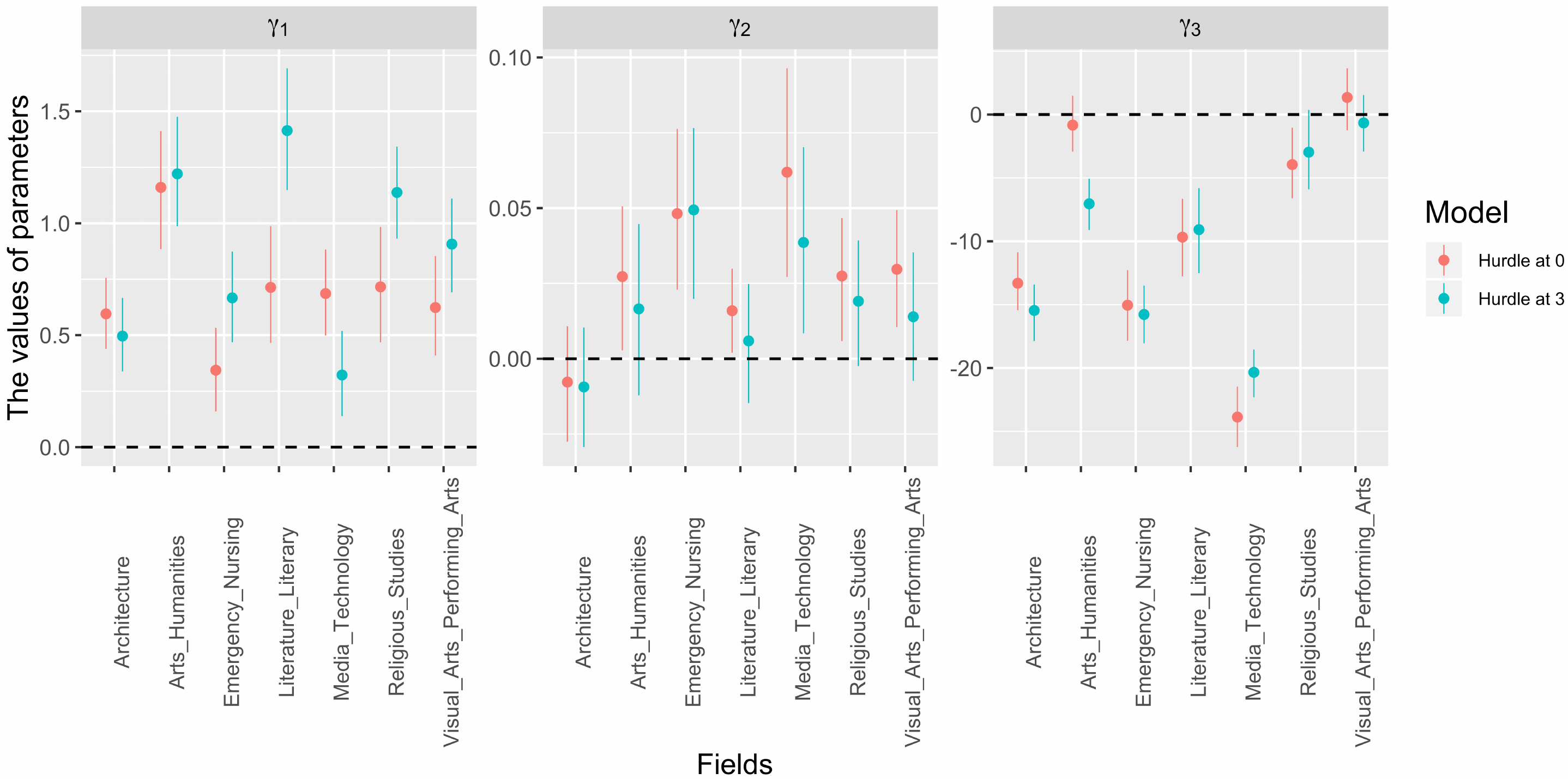}
  \caption{Parameter estimates from the logistic part of the two-part QR models. $\gamma_0$, $\gamma_1$,
$\gamma_2$,$\gamma_3$ are parameters corresponding respectively to intercept, collaboration, title length, and journal internationality in the models}
 
  \label{fig10}
\end{figure}

\newpage
\textbf{Analysing the logistic parts of the two-part models with hurdle at zero and three}\\
The estimates in the logistic part, their credible intervals and standard errors are reported in Table \ref{table:3} and illustrated in Figure\,\ref{fig10}. It is seen that shifting the hurdle from 0 to 3 can influence the significance status and mostly the size of the effects corresponding to the independent variables for the fields with high substantial mass points at 1,2, and 3. For example, for the fields of \textsl{Literature and Literary Theory}, \textsl{Religious Studies}, and \textsl{Visual Arts and Performing Arts}, the absolute effect size of the collaboration gets larger and also, title length ceases to be significant.
In general, in all fields except \textsl{Architecture} and \textsl{Media Technology}, collaboration shows smaller absolute impact size on zero citation in comparison to its impact on low citation (e.g., 0-3 citations). For title length, this trend is inverse over the fields except for \textsl{Emergency Nursing}. The absolute impact size of the longer title on zero citation is slightly larger than on low citation. In addition, for some of the fields, the influence of journal internationality on zero citation is a bit larger than on low citation, but for other fields, it is a bit smaller.

 Collaboration has a significant positive impact on both zero and low cited articles. For example for \textsl{Emergency Nursing} based on the model with hurdle at 0, with greater collaboration, the odds of zero citations increases on average by $41\%$ (more detail: $(\mbox{exp}(0.345)-1)*100=41\%$). In the same field but for the model with hurdle at 3, the odds of low citation increases on average by $96\%$ (more detail: $(\mbox{exp}(0.674)-1)*100=96\%$). Collaboration has its absolute largest impact in the field of \textsl{Arts and Humanities} based on the model with hurdle at 0, while it occurs in the field of \textsl{Literature and Literary Theory} for the model with hurdle at 3.  Based on the previous studies, for example, \citet{Didegah.2014} used the negative binomial hurdle model and showed that collaboration negatively impacts zero citation in most Web of Science subjects.

 Title length in comparison to collaboration and journal internationality has the smallest absolute impact size in all the fields. This effect also shows wider credible intervals for all models with hurdles at 0 and 3 compared to the effects of collaboration and journal internationality. The cases of non-significant status related to this effect are more in comparison to other factors, particularly for the hurdle model at 3. For both hurdle models, the title length has a negative impact on zero citation and also on the low citation for the field of \textsl{Architecture} while for other fields this effect is positive with the small size of the impact. The negativity means the longer title decreases the odds of zero citations based on the model with hurdle at 0 and decreases the odds of low citation in the model with hurdle at 3. In addition, the positivity means the longer title increases zero or low citations. The largest and smallest impact sizes for title length are in \textsl{Media Technology} and \textsl{Architecture} respectively for the model with hurdle at 0, while for the models with hurdle at 3, they are \textsl{Emergency Nursing} and \textsl{Architecture}. \citet{Didegah.2014} showed that title length is a non-significant factor for zero citation for most Web of Science subjects, but for \textsl{Agricultural Sciences}, \textsl{Geosciences}, \textsl{Materials Science}, \textsl{Mathematics}, and \textsl{Physics} title length has a significant positive impact on the odds of zero citation.

Journal internationality has the largest absolute impact on both zero citation and low citation, and also has shorter credible intervals in comparison with collaboration and title length for all models with hurdle at 0 and 3 for all fields. The impact of a greater Gini coefficient (smaller journal internationality) has a significant negative effect on the odds of zero citation and low citation for most of the fields, indicating the direct relationship between journal internationality and zero or low citation. \citet{Didegah.2014} showed that greater journal internationality increases the odds of zero citation for most of the subjects in Web of Science except in \textsl{Space Science} that the effect has a decreasing pattern.

\newpage
\begin{center}
\begin{table}[hhb]{{\extracolsep{\fill}}{}}
\caption{\label{table:3}  Estimates of the vector $\gamma$ and credible intervals and standard deviations from the logistic part of the Bayesian two-part QR model with hurdle at 0, and 3. $\gamma_0$, $\gamma_1$,
$\gamma_2$,$\gamma_3$ are parameters corresponding respectively to intercept, collaboration, title length, and journal internationality in the models
}
\scriptsize{\resizebox{.93\textwidth}{!}{\begin{tabu}{|c|c|c|c|c|c|c|c|c|c|}
\hline
\multirow{2}{*}{Field}&\multirow{2}{*}{Parameters} & \multicolumn{4}{| c |}{ BTPQR with hurdle at 0}& \multicolumn{4}{| c |}{ BTPQR with hurdle at 3} \\ \cline{3-10} 
&& Lower band& Mean & Upper band& Standard deviation& Lower band&  Mean& Upper band&Standard deviation \\ \hline 
\multirow{5}{*}{\textsl{Literature and Literary Theory}}& $\gamma_0$& 6.465 & 8.918&10.875 &1.072 & 4.444& 7.265&10.082 &1.457 \\ 
& $\gamma_1$& 0.460&0.716 &0.992 &0.134 &1.185 &1.485 &1.766 &0.150 \\ 
& $\gamma_2$& 0.001& 0.016& 0.030& 0.008&-0.017 &0.006 &0.026 & 0.011\\ 
& $\gamma_3$& -11.563& -9.527& -7.007& 1.105& -12.844& -9.908&-6.922 &1.511 
\\ \tabucline[.5pt]{1-10}

\multirow{5}{*}{\textsl{ Arts and Humanities}}& $\gamma_0$&-1.703 &0.519& 2.574&1.100&3.821 &5.628&7.635&0.965 \\
& $\gamma_1$&0.884 &1.160&1.412&0.133&0.940 &1.187&1.422&0.123\\
& $\gamma_2$&0.003 &0.027&0.051&0.012&-0.009 &0.019& 0.048&0.015\\
& $\gamma_3$&-2.931 &-0.835& 1.484 &1.137&-10.108 &-8.030&-6.167&1.011
\\  \tabucline[.5pt]{1-10}

\multirow{5}{*}{\textsl{Emergency Nursing}}& $\gamma_0$&11.437 &14.064&16.804&1.399&10.035 &12.087& 14.206&1.081\\
& $\gamma_1$&0.158 &0.345&0.533 &0.095&0.446 &0.674&0.902 &0.115\\
& $\gamma_2$&0.020 &0.048& 0.078 &0.014&0.026 &0.057&0.087&0.015\\
& $\gamma_3$&-18.095 &-15.235& -12.592 &1.425&-17.345 &-15.120&-13.115 &1.094 
\\  \tabucline[.5pt]{1-10}

\multirow{5}{*}{ \textsl{Visual Arts and Performing Arts}}& $\gamma_0$&-3.712 &-1.421& 0.833&1.177&-2.207 &-0.113&2.056&1.133\\
& $\gamma_1$&0.399 &0.622&0.854 &0.115&0.837 &1.056&1.287 &0.111\\
& $\gamma_2$&0.010 &0.029& 0.050&0.010&-0.018 &0.005&0.028&0.010\\
& $\gamma_3$&-0.881 &1.490&3.900 &1.245&-3.822 &-1.506&0.655&1.198 
\\  \tabucline[.5pt]{1-14}

\multirow{5}{*}{\textsl{Architecture}}& $\gamma_0$&10.485 &12.687& 14.943&1.149&11.566 &13.424& 15.209&0.947 \\
& $\gamma_1$&0.433 &0.595& 0.774 &0.088&0.309 &0.472& 0.664 &0.092\\
& $\gamma_2$&-0.027 &-0.009&0.010&0.010&-0.034 &-0.012& 0.012&0.011\\
& $\gamma_3$&-15.367 &-13.068&-10.704 &1.211&-17.418 &-15.538&-13.524 &1.024
\\  \tabucline[.5pt]{1-14}

\multirow{5}{*}{\textsl{Religious Studies}}& $\gamma_0$ &1.500 &4.112&6.813&1.366&-2.269 &0.746& 3.664&1.497\\
& $\gamma_1$&0.457 &0.711&0.963 &0.127&0.984 &1.210&1.410 &0.111\\
& $\gamma_2$&0.004 &0.027& 0.045 &0.011&-0.015 &0.006&0.028&0.011\\
& $\gamma_3$&-6.772 &-3.945&-1.183 &1.423&-5.328 &-2.314&0.809&1.556 
\\  \tabucline[.5pt]{1-10}

\multirow{5}{*}{\textsl{Media Technology}}& $\gamma_0$ &19.359 &21.461& 23.415 &1.033&14.719 &16.389& 17.779&0.791 \\
& $\gamma_1$ &0.499 &0.685&0.867 &0.098&0.099 &0.297&0.488 &0.096\\
& $\gamma_2$ &0.029 &0.062&0.097 &0.017&0.010 &0.043&0.074 &0.016\\
& $\gamma_3$ &-25.749 &-23.635&-21.471 &1.089&-20.710 &-19.177&-17.353&0.867
\\ \hline

\end{tabu}}}
\end{table} 
\end{center}

\newpage
\section{Discussion and conclusion}
Quantile regression enables a deep description of the relationship between independent variables and a dependent variable. It is a useful technique for analysing the entire citation count distribution corresponding to low, moderately, and highly cited articles. Discontinuity and the presence of substantial mass points at lower counts are characteristics of citation counts that make the application of the ``usual''  QR inappropriate. In this research, an update of a Bayesian two-part Hurdle QR model was introduced to scientometrics to address these problems. The original Bayesian two-part hurdle QR model was introduced for the case of count data with a substantial mass point at zero. It allows the zeros and nonzeros data to be modeled separately but simultaneously. For citation count data, as well as a substantial mass point at zero in some fields there can be substantial mass points at lower counts, such as ones, twos, and threes, that influence the estimation of the model. Therefore, we introduce a method to shift the hurdle forward to discard the effect of the substantial mass points on the estimation of the model for fields with many low cited articles. Articles without more citations than the hurdle are regarded as ``low cited articles''. In this new update, the model enables analyses of the citation counts of low cited articles simultaneously but separately from those of the moderately and highly cited articles.It uses jittering for the citation counts greater than the hurdle to render such data continuous. The model benefits from the power of its QR portion for modeling the different quantiles of the jittered citation counts, and its logistic portion for analysing the influence of factors such as collaboration, title length, and journal internationality on the chances of an article receiving few citations. The usefulness and applicability of the method were illustrated based on both simulated and real citation count data. The simulation showed that the QR part of the two-part model with a hurdle point past the substantial mass points in the data, gives  more accurate estimates at quantiles just beyond the hurdle based on the indicator of the mean squared error of the estimates of the coefficients corresponding to the independent variables in the model. Moreover, the QR part of the two-part QR models provides smaller prediction errors at the cost of slightly wider credible intervals for the parameter estimates in comparison to the Bayesian QR. Citation data from seven Scopus fields were also considered and three models including Bayesian QR, Bayesian two-part QR with hurdle at 0, and Bayesian two-part QR with hurdle at 3 were fitted to the data. The results of the Bayesian QR model based on the whole data shows a pattern with more fluctuations for the independent variables over the quantiles. However, the two-part models with hurdle at 0 and 3, generally show a smoother trend of the estimates over the quantiles for most of the fields. Shifting the hurdle from 0 to a larger point and passing the substantial mass points in the data influence the impact size, the significance status, and the width of the credible intervals illustrating the importance of choosing the hurdle appropriately.

In summary, we have shown that the proposed hurdle-at-three model has many advantages over the hurdle-at-zero model of \cite{King.et.al2019} for the modelling of citation count data for fields with large percentages of articles with few citations.

\section{Acknowledgments}
The authors would like to thank Clay King for his helpful comments.

\section{Author contributions}
Marzieh Shahmandi: Data curation, Investigation, Methodology, Writing—original draft, Validation, Software. Paul Wilson: Supervision, Writing—review $\&$ editing. Mike Thelwall: Supervision, Writing—review $\&$ editing.

\section{Competing interests}
No competing interests.

\section{Funding information}
No funding was received for this study.

\section{Data availability}
The processed data used to produce the tables and figures for the section of citation count example are available in the supplementary material (https://doi.org/10.6084/m9.figshare.14742939.v4).


\end{document}